\documentclass{srg1col}

\title{Heat and Moisture Budget Analysis \\[0.5ex] 
       with an Improved Form of Moist Thermodynamics}

\authors{A}{Wayne H. Schubert, Paul E. Ciesielski, and Richard H. Johnson}
\affiliation{A}{Department of Atmospheric Science, Colorado State University,
              Fort Collins, Colorado, USA}
\corresponding{Paul E. Ciesielski}{paulc@atmos.colostate.edu}
\docinfo{25 October 2018}

\abstract{In order to understand the effects of cumulus convection on
large-scale atmospheric motions, heat and moisture budget analyses are often
performed using data from an array of radiosonde stations. Ever since the
pioneering work of Yanai et al.~(1973), such budgets have been based on
approximate forms of moist thermodynamics. This paper presents an improved
form of moist thermodynamics for such budget studies.}

\begin{document}
\maketitle

\section{Introduction}

     Ever since the pioneering work of Yanai et al.~(1973), heat and 
moisture budget studies, and their interpretation in terms of the convective 
flux of moist static energy, have been performed with approximate forms of 
moist thermodynamics. For example, such studies often begin with the 
assumption that moist static energy $h=c_{pa} T + gz + Lq$, defined with 
the latent heat of condensation $L$ assumed constant, is materially conserved 
(except for radiative effects). This misses important physical effects of ice, 
such as the formation of stable layers near the melting level. This is only 
one of several consequences 
of using this approximate form of moist thermodynamics. While the effects 
of approximate moist thermodynamics are generally small at a given time, 
their accumulated effects over longer periods can be substantial. 
This was apparent to Johnson and Ciesielski (2000) in computing long-term 
averaged budget residuals such as precipitation and column average radiation 
for TOGA COARE. 

    Since its introduction in the late 1990s, the constrained variational analysis 
(CVA) approach to atmospheric budgets has proved to be an invaluable tool for 
computing accurate large-scale forcing fields from sounding data networks 
(Zhang and Lin 1997). In this method the atmospheric state variables, typically 
derived from sounding data, are adjusted in a minimal way comparable to measurement 
uncertainty to conserve vertical constraints of mass, moisture, static energy and 
momentum. In so doing, CVA produces fields that are consistent with these constraints. 
These additional constraints result in significant improvements in budget diagnostics 
especially during periods when the original sounding data are missing or of questionable 
quality. While the effects of measurement and sampling errors in sounding data 
(Mapes et al.~2003) are effectively minimized with CVA, accurate large-scale 
diagnostics in this method are dependent on reliable constraints, particularly 
rainfall (Xie et al.~2004). 

     Ooyama (1990, 2001) proposed a very accurate form of moist thermodynamics 
for use in tropical models.  This accurate 
formulation of moist thermodynamics is not limited to modeling 
studies but can also be used in heat and moisture budget studies. 
With the advent of GPS sondes and the recent availability of certain satellite  
data products, this more accurate treatment of moist thermodynamics provides
opportunities to refine our understanding of tropical systems. This paper develops 
more accurate methods, which can utilize ancillary ground-based measurements 
and satellite data products, for the diagnosis of heat and moisture budgets. 
The overall goal of this diagnostic technique development is to provide 
the observational analysis community with an improved method for computing
atmospheric budgets, which has the additional advantage of having a 
complementary numerical modeling framework with the same moist thermodynamics. 
This should allow better comparisons between satellite latent heating products 
and numerical model simulations. 

\section{Review of the standard theoretical basis for heat and moisture budget studies}

     In the standard theory (Yanai et al.~1973), mass conservation 
relations are written for only two forms of matter: dry air and 
water vapor. In addition, a thermodynamic relation 
for the dry static energy is included. Let $\rho_a$ denote the mass density 
of dry air,\footnote{A list of key symbols for the nonhydrostatic model is
given in appendix A at the end of this paper.} 
$\rho_v$ the mass density of water vapor, and $s=c_{pa}T+gz$ the dry static energy.   
The flux forms of the prognostic equations for $\rho_a$, $s$, and $\rho_v$ are given 
by\footnote{Although the standard theory is usually derived using pressure as the 
vertical coordinate, here we use the height coordinate, which allows direct 
comparison with the more accurate theory presented in section 3.} 
\begin{equation}                                
       \frac{\partial \rho_a   }{\partial t} 
     + \frac{\partial(\rho_a u)}{\partial x}  
     + \frac{\partial(\rho_a v)}{\partial y}
     + \frac{\partial(\rho_a w)}{\partial z} = 0,      
\label{eq2.1}
\end{equation}
\begin{equation}                                
       \frac{\partial(\rho_a s) }{\partial t} 
     + \frac{\partial(\rho_a us)}{\partial x}  
     + \frac{\partial(\rho_a vs)}{\partial y}
     + \frac{\partial(\rho_a ws)}{\partial z} 
     = L\rho_a (c - e) + \rho_a Q_R,            
\label{eq2.2}
\end{equation}
\begin{equation}                                
       \frac{\partial \rho_v   }{\partial t} 
     + \frac{\partial(\rho_v u)}{\partial x}  
     + \frac{\partial(\rho_v v)}{\partial y}
     + \frac{\partial(\rho_v w)}{\partial z} = -\rho_a (c - e),   
\label{eq2.3}
\end{equation}
where $c$ is the rate of condensation, $e$ the rate of evaporation, and $Q_R$ the 
radiative heating rate. In the 
standard theory, the latent heat of condensation $L$ is usually taken to be a 
constant, in contrast to the more accurate treatment discussed in section 3. 
Instead of working with the mass density of water vapor, $\rho_v$, the standard theory 
usually works with the water vapor mixing ratio $q_v$, which is related to the water 
vapor mass density by $\rho_v=\rho_a q_v$. The equation for $q_v$ is obtained by simply 
replacing $\rho_v$ by $\rho_a q_v$ everywhere in (\ref{eq2.3}). 

    Taking the horizontal average of (\ref{eq2.1})--(\ref{eq2.3}), we obtain 
\begin{equation}                               
       \frac{\partial\bar{\rho}_a}{\partial t}  
     + \frac{\partial(\overline{\rho_a u})}{\partial x}  
     + \frac{\partial(\overline{\rho_a v})}{\partial y}
     + \frac{\partial(\overline{\rho_a w})}{\partial z} = 0,   
\label{eq2.4}
\end{equation}
\begin{equation}                               
       \frac{\partial(\overline{\rho_a   s})}{\partial t}    
     + \frac{\partial(\overline{\rho_a u s})}{\partial x}  
     + \frac{\partial(\overline{\rho_a v s})}{\partial y}
     + \frac{\partial(\overline{\rho_a w s})}{\partial z} 
     = L\overline{\rho_a (c - e)} + \overline{\rho_a Q_R},  
\label{eq2.5}
\end{equation}
\begin{equation}                               
       \frac{\partial(\overline{\rho_a   q_v})}{\partial t} 
     + \frac{\partial(\overline{\rho_a u q_v})}{\partial x}  
     + \frac{\partial(\overline{\rho_a v q_v})}{\partial y}
     + \frac{\partial(\overline{\rho_a w q_v})}{\partial z} 
     = -\overline{\rho_a (c - e)}. 
\label{eq2.6}
\end{equation}
Making the usual approximations such as neglecting horizontal eddy flux terms 
involving $\overline{u's'}$, $\overline{v's'}$, $\overline{u'q_v'}$, 
and $\overline{v'q_v'}$, and then making the approximations 
$\overline{\rho_a w s}\approx \bar{\rho}_a \bar{w}\bar{s} 
                            + \bar{\rho}_a \overline{w's'}$, and  
$\overline{\rho_a wq_v}\approx \bar{\rho}_a \bar{w}\bar{q}_v 
                             + \bar{\rho}_a \overline{w'q_v'}$,  
we can simplify (\ref{eq2.4})--(\ref{eq2.6}) to  
\begin{equation}                                 
           \frac{\partial \bar{\rho}_a         }{\partial t} 
         + \frac{\partial(\bar{\rho}_a \bar{u})}{\partial x}  
         + \frac{\partial(\bar{\rho}_a \bar{v})}{\partial y}
         + \frac{\partial(\bar{\rho}_a \bar{w})}{\partial z} = 0,  
\label{eq2.7}
\end{equation}
\begin{equation}                               
       \frac{\partial(\bar{\rho}_a         \bar{s})}{\partial t}    
     + \frac{\partial(\bar{\rho}_a \bar{u} \bar{s})}{\partial x}  
     + \frac{\partial(\bar{\rho}_a \bar{v} \bar{s})}{\partial y}
     + \frac{\partial(\bar{\rho}_a \bar{w} \bar{s})}{\partial z} 
     =-\frac{\partial(\bar{\rho}_a \overline{w's'})}{\partial z}
     + L\bar{\rho}_a (\bar{c} - \bar{e}) + \bar{\rho}_a \bar{Q}_R,  
\label{eq2.8}
\end{equation}
\begin{equation}                               
       \frac{\partial(\bar{\rho}_a         \bar{q}_v)}{\partial t} 
     + \frac{\partial(\bar{\rho}_a \bar{u} \bar{q}_v)}{\partial x}  
     + \frac{\partial(\bar{\rho}_a \bar{v} \bar{q}_v)}{\partial y}
     + \frac{\partial(\bar{\rho}_a \bar{w} \bar{q}_v)}{\partial z} 
     =-\frac{\partial(\bar{\rho}_a \overline{w'q_v'})}{\partial z} 
     - \bar{\rho}_a (\bar{c} - \bar{e}).   
\label{eq2.9}
\end{equation}
Using (\ref{eq2.7}), the flux forms (\ref{eq2.8}) and (\ref{eq2.9}) can be
converted to the advective forms 
\begin{equation}                                 
     Q_1 \equiv   \frac{\partial\bar{s}}{\partial t} 
         + \bar{u}\frac{\partial\bar{s}}{\partial x}  
         + \bar{v}\frac{\partial\bar{s}}{\partial y}
         + \bar{w}\frac{\partial\bar{s}}{\partial z} 
         =-\frac{\partial(\bar{\rho}_a \overline{w's'})}{\bar{\rho}_a\partial z} 
         + L(\bar{c} - \bar{e}) + \bar{Q}_R,                 
\label{eq2.10}
\end{equation}
\begin{equation}                                 
     Q_2 \equiv -L\left(\frac{\partial\bar{q}_v}{\partial t} 
               + \bar{u}\frac{\partial\bar{q}_v}{\partial x}  
               + \bar{v}\frac{\partial\bar{q}_v}{\partial y}
               + \bar{w}\frac{\partial\bar{q}_v}{\partial z}\right) 
               = L\frac{\partial(\bar{\rho}_a \overline{w'q_v'})}{\bar{\rho}_a\partial z} 
               + L(\bar{c} - \bar{e}),    
\label{eq2.11}
\end{equation}
with the first equality in (\ref{eq2.10}) serving as the definition of the apparent 
heat source $Q_1$ and the first equality in (\ref{eq2.11}) serving as the definition 
of the apparent moisture sink $Q_2$. 
If a network of upper air observations gives us $\bar{u}$, $\bar{v}$, 
$\bar{\rho}_a$, $\bar{s}$, and $\bar{q}_v$, and if we compute $\bar{w}$ 
from (\ref{eq2.7}), then we can compute the apparent heat source $Q_1$ from the 
large-scale terms in (\ref{eq2.10}) and the apparent moisture sink $Q_2$ from the 
large-scale terms in (\ref{eq2.11}). Subtracting (\ref{eq2.11}) from (\ref{eq2.10}), we obtain 
\begin{equation}                                 
               \frac{\partial\bar{h}}{\partial t} 
      + \bar{u}\frac{\partial\bar{h}}{\partial x}  
      + \bar{v}\frac{\partial\bar{h}}{\partial y}
      + \bar{w}\frac{\partial\bar{h}}{\partial z} - \bar{Q}_R
      = Q_1 - Q_2 - \bar{Q}_R  
      =-\frac{\partial(\bar{\rho}_a\overline{w'h'})}{\bar{\rho}_a\,\partial z},
\label{eq2.12}         
\end{equation}
where $\bar{\rho}_a\overline{w'h'}$ is the vertical eddy flux of moist static energy, 
with $h$ defined by $h=c_{pa}T+gz+Lq=s+Lq_v$. 
Assuming that $\overline{w'h'}$ vanishes at the tropopause height $z=z_T$, integration 
of (\ref{eq2.12}) yields 
\begin{equation}                                 
      \bar{\rho}_a\overline{w'h'} 
    = \int_{z}^{z_T} \left(\frac{\partial\bar{h}}{\partial t} 
                  + \bar{u}\frac{\partial\bar{h}}{\partial x}  
                  + \bar{v}\frac{\partial\bar{h}}{\partial y}
                  + \bar{w}\frac{\partial\bar{h}}{\partial z} - \bar{Q}_R \right) \bar{\rho}_a(z') \, dz'
    = \int_{z}^{z_T} \left(Q_1 - Q_2 - \bar{Q}_R\right) \bar{\rho}_a(z') \, dz'.  
\label{eq2.13}
\end{equation}           
Once the entire vertical profile of the convective flux $\bar{\rho}_a\overline{w'h'}$ 
has been found from (\ref{eq2.13}), its surface value can be checked against independent surface 
flux measurements (e.g., Nitta and Esbensen 1974). Cloud mass fluxes can then be 
deduced by interpreting the convective 
flux in terms of a bulk cloud model or a spectral cloud model, as discussed in the 
review article by Yanai and Johnson (1993). 

     Another useful integral constraint can be obtained from (\ref{eq2.11}). Assuming that 
$\overline{w'q_v'}$ vanishes at the tropopause height $z=z_T$, integration of 
(\ref{eq2.11}) leads to 
\begin{equation}                                 
   P_o = E_o - \int_{o}^{z_T} \left(\frac{\partial\bar{q}_v}{\partial t}
         + \bar{u}\frac{\partial\bar{q}_v}{\partial x}
         + \bar{v}\frac{\partial\bar{q}_v}{\partial y}
         + \bar{w}\frac{\partial\bar{q}_v}{\partial z} \right)
           \bar{\rho}_a(z) \, dz,  
\label{eq2.14}
\end{equation}
where $E_o=\left(\bar{\rho}_a\overline{w'q_v'}\right)_o$ is the surface evaporation 
and $P_o=\int_0^{z_T}(\bar{c}-\bar{e})\bar{\rho}_a\, dz$ is the precipitation. Thus, 
with an estimate of $E_o$ from surface data, precipitation can be diagnosed as a budget
residual. In a similar fashion, (\ref{eq2.13}) can be rearranged to give a relationship
for the column-net radiation 
\begin{equation}                                   
   \langle\bar{Q}_R\rangle = \langle Q_1\rangle - \langle Q_2\rangle - S_o - LE_o,
\label{eq2.15}
\end{equation}
where $\langle(\cdot)\rangle = \int_{o}^{z_T} \left(\cdot\right) \bar{\rho}_a(z) \, dz$ 
and $S_o$ is the surface sensible heat flux. In past studies, these budget-derived 
vertically integrated quantities in (\ref{eq2.14}) and (\ref{eq2.15}) have been compared 
to independent estimates to gain insights into the quality of the budget analyses 
(e.g., Johnson and Ciesielski 2000, Johnson et al.~2015). 

     In summary, the theoretical basis for the standard heat and moisture budget 
analysis is (\ref{eq2.1})--(\ref{eq2.3}). This theoretical basis fits nicely with what can be measured  
with a radiosonde sounding array: vertical profiles of the dry static energy $\bar{s}$, 
the water vapor mixing ratio $\bar{q}_v$, and the horizontal wind components $\bar{u}$ 
and $\bar{v}$. Although radiosondes cannot directly measure the vertical component 
$\bar{w}$, the continuity equation (\ref{eq2.7}) can be used to compute the vertical component 
if the horizontal components are observed with sufficient accuracy. An important 
breakthrough in this regard has been the application of GPS technology to accurately  
determine balloon positions as a function of height and hence the horizontal wind 
components. A nagging issue in many field programs has been the quality of the $q_v$ 
observations from radiosondes. However, recently there have been significant 
advances with the development of small sensors constructed from porous polymers, 
whose capacitance varies with humidity. When equipped with a small heating element, 
these sensors can be kept free of ice and thereby yield accurate humidity measurements 
over the entire troposphere.\footnote{The Vaisala RS41 sonde is an example of this 
improved technology.}  Thus advancements in radiosonde technology have greatly reduced
measurement errors in the basic atmospheric fields which plagued earlier budget studies (e.g.,
GATE and TOGA COARE). 

     A second source of uncertainty in budget analyses based on networks of
sounding stations is related to data sampling. Sampling errors come both 
from systematic errors associated with the network geometry (Ciesielski et al.~1999) 
and from random errors of representativeness (Mapes et al.~2003). An example of 
the former error was discussed by Katsumata et al.~(2011), who found that a  
triangular sounding network was unable to properly capture the divergence signal 
associated with the Rossby and inertia-gravity wave components of an equatorial flow. 
The latter errors result from using point measurements from radiosondes and assuming 
that they are representative of a space-time region comparable to the distance between 
soundings. In reality the space scales between stations and timescales between sounding 
releases contain a wide variety of unresolved circulations and thermodynamic structures. 
To reduce the impact of random sampling errors in budget calculations, analyses are typically 
presented as spatial and temporal averages, which reduces such errors in a way that is 
roughly consistent with elementary $N^{-1/2}$ sampling statistics, where $N$ is the number 
of observation times. For example, sampling 
errors would be halved for 1-day averages (assuming 6-h analyses) and halved again for 
4-day averages. By adjusting the state variables to be consistent with vertical constraints, 
measurement and sampling errors are effectively minimized in the CVA budget approach 
(Zhang and Lin 1997). Analysis uncertainties with the CVA method arise rather from sampling 
errors in the vertical constraint fields (e.g., rainfall and column-net radiation).  

     In spite of past radiosonde limitations and network sampling errors, the application of the standard 
budget method to tropical and subtropical observations has been extremely successful 
(Yanai et al.~1973, Ogura and Cho 1973, Nitta and Esbensen 1974, Nitta 1975, 
Johnson 1980, Johnson and Young 1983, Johnson 1984, Gallus and Johnson 1991, 
Yanai and Johnson 1993, Ciesielski et al.~1999, Zhang et al.~2001, 
Schumacher et al.~2007, 2008, Johnson et al.~2016) and has provided 
a basis for many of the cumulus parameterizations used in numerical weather 
prediction models and in general circulation models.  The method 
is particularly well-suited for models which predict only the large-scale water
vapor field (and not the airborne condensate) and do not keep track of the details 
of precipitation. However, large-scale models are becoming more complete in their 
physics, so it is important to discuss the theoretical shortcomings of the standard approach. 
Three obvious ones are as follows. 
\begin{itemize} 
\item The moisture budget is based on water vapor only, thereby neglecting the 
      advective and storage effects of cloud and precipitating condensate. These 
      effects can be significant, especially in situations with large variability 
      in fractional 
      cloudiness.\footnote{Discussions of cloud water and rainwater storage 
      effects are given by McNab and Betts (1978) and Gallus and Johnson (1991) 
      for convective cases over land and by Johnson (1980) for oceanic (GATE) cases. 
      Rainwater storage effects can be significant when precipitation production has 
      ceased but existing rainwater continues to fall to the earth's surface. Horizontal 
      advective effects on condensed water can be important in the trailing stratiform 
      regions of tropical and midlatitude squall lines.}
      This shortcoming can be corrected by generalizing the water 
      vapor budget equation to a budget equation for the total airborne moisture 
      (vapor and cloud condensate) and by including a separate budget equation 
      for precipitating condensate (water and/or ice). See (\ref{eq3.2}) and (\ref{eq3.3}) below. 
\item Because $L$ is usually taken as a constant, ice effects are not 
      included.\footnote{Heat and moisture budget studies that include ice effects 
      have been performed by Johnson and Young (1983) for tropical mesoscale anvil 
      clouds and by Gallus and Johnson (1991) for an intense midlatitude squall 
      line.} A variable $L$ should be included to capture the latent heat of fusion. 
      Also, a slow terminal velocity for ice and a faster terminal velocity 
      for rain below the melting zone should be used. This could lead to more 
      accurate diagnosis of melting layer inversions in tropical regions 
      (Johnson et al.~1996). 
\item The thermodynamic principle is approximate since it is based on either dry 
      static energy $s$ or moist static energy $h$. A more accurate principle 
      can be developed based on the entropy density for a sample volume that 
      contains contributions from dry air, water vapor, cloud condensate, and 
      precipitation. See (\ref{eq3.4}) below. 
\end{itemize}

     In the next section we outline the basic framework for developing a more 
precise budget analysis technique which overcomes the shortcomings in the theoretical foundation
of the standard approach. This new technique makes use of a more accurate model of the moist 
atmosphere, but it demands more data than radiosondes alone can provide. Thus, 
it must be used in conjunction with recently available satellite data products 
such as TRMM and GPM.

\section{A more accurate model of the moist atmosphere}     

     Now consider atmospheric matter to consist of dry air, airborne moisture (vapor 
and cloud condensate), and precipitation. Let $\rho_a$ denote the mass density 
of dry air, $\rho_m=\rho_v+\rho_c$ the mass density of airborne moisture (consisting 
of the sum of the mass densities of water vapor $\rho_v$ and airborne condensed water 
$\rho_c$), and $\rho_r$ the mass density of precipitating water substance.\footnote{Above 
the freezing level, the precipitating water substance is assumed to be ice, while below 
the freezing level, it is assumed to be liquid. The smoothness of the transition between 
liquid and ice is adjustable, as explained in Fig.~1.} The total 
mass density $\rho$ is given by $\rho=\rho_a+\rho_v+\rho_c+\rho_r=\rho_a+\rho_m+\rho_r$.  
The flux forms of the prognostic equations for $\rho_a$, $\rho_m$, and $\rho_r$ are 
given by\footnote{In modeling the cloudy, precipitating atmosphere, two conceptual 
errors can easily be made. The first is that, since moisture occurs in the three forms 
$\rho_v,\rho_c,\rho_r$, then three separate prognostic equations for these fields are 
required in the analysis. However, only prognostic equations for $\rho_m$ and $\rho_r$ 
are needed since $\rho_m$ is entirely vapor (i.e., $\rho_v=\rho_m$ and $\rho_c=0$) in 
subsaturated conditions, while $\rho_v=\rho_v^*(T)$ and $\rho_c=\rho_m-\rho_v^*(T)$ in 
saturated conditions. In other words, the partition of the predicted variable $\rho_m$ 
into its components $\rho_v$ and $\rho_c$ is done diagnostically, as discussed in 
section 4. The second conceptual error is that $\rho_c$ and $\rho_r$ should be grouped 
together as a single prognostic field, since both involve condensate. However, this 
leads to considerable difficulty since $\rho_c$ and $\rho_r$ move at different 
vertical velocities. Thus, the number of prognostic equations required to model a 
cloudy, precipitating atmosphere is seven (for $u,v,w,\rho_a,\rho_m,\rho_r,\sigma$), 
while the number required to model a dry atmosphere is five (for $u,v,w,\rho_a,\sigma_a$).} 
\begin{equation}                               
       \frac{\partial \rho_a   }{\partial t}  
     + \frac{\partial(\rho_a u)}{\partial x}  
     + \frac{\partial(\rho_a v)}{\partial y}
     + \frac{\partial(\rho_a w)}{\partial z} = 0,   
\label{eq3.1}
\end{equation}
\begin{equation}                               
       \frac{\partial \rho_m   }{\partial t} 
     + \frac{\partial(\rho_m u)}{\partial x}  
     + \frac{\partial(\rho_m v)}{\partial y}
     + \frac{\partial(\rho_m w)}{\partial z} = -Q_r, 
\label{eq3.2}
\end{equation}
\begin{equation}                               
       \frac{\partial \rho_r   }{\partial t}    
     + \frac{\partial(\rho_r u)}{\partial x}  
     + \frac{\partial(\rho_r v)}{\partial y}
     + \frac{\partial(\rho_r w_r)}{\partial z} = Q_r. 
\label{eq3.3}
\end{equation}
Note that (\ref{eq3.1})--(\ref{eq3.3}) generalize (\ref{eq2.1}) and (\ref{eq2.3}) 
in two regards: they include a budget equation for precipitating water 
substance\footnote{The treatment presented here can be considered a bulk 
microphysics scheme, since the entire precipitation field is described by the 
single dependent variable $\rho_r(x,y,z,t)$. For an interesting discussion of 
the treatment of hydrometeor sedimentation in bulk and hybrid bulk-bin 
microphysics schemes, see Morrison (2012).} and an equation for total airborne
water (vapor plus cloud) rather than vapor alone.  Also note that 
these three mass continuity equations contain two vertical velocities, 
$w$ and $w_r$, where $w$ denotes the vertical velocity  
of dry air and airborne moisture, and $w_r$ denotes the vertical velocity of 
the precipitating water substance, so that $w_\infty=w_r-w$ is the terminal 
velocity of the precipitating water substance relative to the dry air  
and airborne moisture. The term $Q_r$, on the right hand
sides of (\ref{eq3.2}) and (\ref{eq3.3}), is the rate of 
conversion from airborne moisture to precipitation; this term can be 
positive (e.g., the collection of cloud droplets by rain) or negative (e.g., 
the evaporation of precipitation falling through unsaturated air). 
 
     To generalize (\ref{eq2.2}) we consider the total entropy density of moist air. 
The total entropy density is $\sigma=\sigma_a+\sigma_m+\sigma_r$, consisting 
of the sum of the entropy densities of dry air, airborne moisture, and 
precipitation.\footnote{Note that it is the entropy densities 
($\sigma_a,\sigma_m,\sigma_r$), not the specific entropies ($s_a,s_m,s_r$), 
that are additive.}  Since the vertical entropy flux is given by 
$\sigma_a w+\sigma_m w+\sigma_r w_r=\sigma w+\sigma_r w_\infty$, we can write 
the flux form of the entropy conservation principle as  
\begin{equation}                           
       \frac{\partial \sigma   }{\partial t} 
     + \frac{\partial(\sigma u)}{\partial x}  
     + \frac{\partial(\sigma v)}{\partial y}
     + \frac{\partial(\sigma w + \sigma_r w_\infty)}{\partial z} = Q_\sigma, 
\label{eq3.4}
\end{equation}
where $Q_\sigma$ denotes nonconservative processes such as radiation.  For 
the details on how $\sigma$ is computed, see Ooyama (2001) and the brief 
discussion in section 4. Note that the vertical entropy flux by precipitation 
(which is assumed to fall at the wet-bulb temperature) is incorporated
through the $\sigma_r(w+w_\infty)$ terms in (\ref{eq3.4}). 

     Now define the mixing ratio of airborne moisture by $q_m=\rho_m/\rho_a$, 
the mixing ratio of precipitation by $q_r=\rho_r/\rho_a$, the 
dry-air-specific\footnote{The term ``dry-air-specific" is used because 
$\sigma$ is divided by the dry air density $\rho_a$, instead of the total 
density $\rho$, i.e., the specific entropy ${\mathcal S}$ of moist air is 
measured per unit mass of dry air.} entropy of 
moist air by ${\mathcal S}=\sigma/\rho_a$, and the dry-air-specific entropy 
of precipitation by ${\mathcal S}_r=\sigma_r/\rho_a$, so that we can replace 
$\rho_m$ by $\rho_a q_m$ in (\ref{eq3.2}), $\rho_r$ by $\rho_a q_r$ in (\ref{eq3.3}), 
$\sigma$ by $\rho_a {\mathcal S}$ and $\sigma_r$ by $\rho_a {\mathcal S}_r$ in (\ref{eq3.4}). 
Then, taking the horizontal average of (\ref{eq3.1})--(\ref{eq3.4}), we obtain 
\begin{equation}                               
       \frac{\partial\bar{\rho}_a}{\partial t}  
     + \frac{\partial(\overline{\rho_a u})}{\partial x}  
     + \frac{\partial(\overline{\rho_a v})}{\partial y}
     + \frac{\partial(\overline{\rho_a w})}{\partial z} = 0,   
\label{eq3.5}
\end{equation}
\begin{equation}                               
       \frac{\partial(\overline{\rho_a   q_m})}{\partial t} 
     + \frac{\partial(\overline{\rho_a u q_m})}{\partial x}  
     + \frac{\partial(\overline{\rho_a v q_m})}{\partial y}
     + \frac{\partial(\overline{\rho_a w q_m})}{\partial z} 
     = -\bar{Q}_r, 
\label{eq3.6}
\end{equation}
\begin{equation}                               
       \frac{\partial(\overline{\rho_a   q_r})}{\partial t}    
     + \frac{\partial(\overline{\rho_a u q_r})}{\partial x}  
     + \frac{\partial(\overline{\rho_a v q_r})}{\partial y}
     + \frac{\partial(\overline{\rho_a w q_r}
                    + \overline{\rho_a w_\infty q_r})}{\partial z} 
     = \bar{Q}_r. 
\label{eq3.7}
\end{equation}
\begin{equation}                               
       \frac{\partial(\overline{\rho_a  {\mathcal S}})}{\partial t}    
     + \frac{\partial(\overline{\rho_a u{\mathcal S}})}{\partial x}  
     + \frac{\partial(\overline{\rho_a v{\mathcal S}})}{\partial y}
     + \frac{\partial(\overline{\rho_a w{\mathcal S}} 
                    + \overline{\rho_a w_\infty{\mathcal S}_r})}{\partial z} 
     = \bar{Q}_\sigma. 
\label{eq3.8}
\end{equation}
Now approximate (\ref{eq3.5})--(\ref{eq3.8}) by 
\begin{equation}                               
       \frac{\partial \bar{\rho}_a         }{\partial t}  
     + \frac{\partial(\bar{\rho}_a \bar{u})}{\partial x}  
     + \frac{\partial(\bar{\rho}_a \bar{v})}{\partial y}
     + \frac{\partial(\bar{\rho}_a \bar{w})}{\partial z} = 0,   
\label{eq3.9}
\end{equation}
\begin{equation}                               
       \frac{\partial(\bar{\rho}_a        \bar{q}_m)}{\partial t} 
     + \frac{\partial(\bar{\rho}_a \bar{u}\bar{q}_m)}{\partial x}  
     + \frac{\partial(\bar{\rho}_a \bar{v}\bar{q}_m)}{\partial y}
     + \frac{\partial(\bar{\rho}_a \bar{w}\bar{q}_m)}{\partial z} 
     =-\frac{\partial(\bar{\rho}_a \overline{w'q'_m})}{\partial z} 
     - \bar{Q}_r, 
\label{eq3.10}
\end{equation}
\begin{equation}                               
       \frac{\partial(\bar{\rho}_a        \bar{q}_r)}{\partial t}    
     + \frac{\partial(\bar{\rho}_a \bar{u}\bar{q}_r)}{\partial x}  
     + \frac{\partial(\bar{\rho}_a \bar{v}\bar{q}_r)}{\partial y}
     + \frac{\partial(\bar{\rho}_a \bar{w}\bar{q}_r)}{\partial z}
     =-\frac{\partial(\bar{\rho}_a \overline{w'q_r'} 
                    + \bar{\rho}_a \overline{w_\infty q_r})}{\partial z} 
     + \bar{Q}_r. 
\label{eq3.11}
\end{equation}
\begin{equation}                               
       \frac{\partial(\bar{\rho}_a        \bar{\mathcal S})}{\partial t}    
     + \frac{\partial(\bar{\rho}_a \bar{u}\bar{\mathcal S})}{\partial x}  
     + \frac{\partial(\bar{\rho}_a \bar{v}\bar{\mathcal S})}{\partial y}
     + \frac{\partial(\bar{\rho}_a \bar{w}\bar{\mathcal S})}{\partial z}
     =-\frac{\partial(\bar{\rho}_a\overline{w'{\mathcal S}'} 
                    + \bar{\rho}_a\overline{w_\infty{\mathcal S}_r})}{\partial z} 
     + \bar{Q}_\sigma. 
\label{eq3.12}
\end{equation}
It is interesting to note that the sum of (\ref{eq3.9}) through (\ref{eq3.11}) yields the 
prognostic equation for the total density. When this equation for the total 
density is integrated over the entire domain, we observe that the total mass 
in the domain is not conserved (even though $\bar{w}=0$ at the lower boundary) 
because of evaporation and precipitation at the lower boundary.  
Using (\ref{eq3.9}), the three flux forms (\ref{eq3.10})--(\ref{eq3.12}) can be converted to the 
advective forms 
\begin{equation}                                  
                  \frac{\partial\bar{q}_m}{\partial t} 
         + \bar{u}\frac{\partial\bar{q}_m}{\partial x}  
         + \bar{v}\frac{\partial\bar{q}_m}{\partial y}
         + \bar{w}\frac{\partial\bar{q}_m}{\partial z} 
         =-\frac{\partial(\bar{\rho}_a \overline{w'q'_m})}{\bar{\rho}_a\,\partial z} 
         - \frac{\bar{Q}_r}{\bar{\rho}_a}.   
\label{eq3.13}
\end{equation}
\begin{equation}                                  
                  \frac{\partial\bar{q}_r}{\partial t} 
         + \bar{u}\frac{\partial\bar{q}_r}{\partial x}  
         + \bar{v}\frac{\partial\bar{q}_r}{\partial y}
         + \bar{w}\frac{\partial\bar{q}_r}{\partial z} 
         =-\frac{\partial(\bar{\rho}_a \overline{w'q'_r}
                        + \bar{\rho}_a \overline{w_\infty q_r})} 
                                  {\bar{\rho}_a\,\partial z} 
         + \frac{\bar{Q}_r}{\bar{\rho}_a}.       
\label{eq3.14}
\end{equation}
\begin{equation}                                  
                  \frac{\partial\bar{\mathcal S}}{\partial t} 
         + \bar{u}\frac{\partial\bar{\mathcal S}}{\partial x}  
         + \bar{v}\frac{\partial\bar{\mathcal S}}{\partial y}
         + \bar{w}\frac{\partial\bar{\mathcal S}}{\partial z} 
         =-\frac{\partial(\bar{\rho}_a \overline{w'{\mathcal S}'}
                        + \bar{\rho}_a \overline{w_\infty {\mathcal S}_r})}
                                  {\bar{\rho}_a\,\partial z} 
         + \frac{\bar{Q}_\sigma}{\bar{\rho}_a}.      
\label{eq3.15}
\end{equation}
Equations (\ref{eq3.9}) and (\ref{eq3.13})--(\ref{eq3.15}) form the basis for budget studies with the 
improved form of moist thermodynamics. 

     By assuming that 
$\bar{\rho}_a\overline{w'{\mathcal S}'}+\bar{\rho}_a\overline{w_\infty {\mathcal S}_r}$ 
vanishes at the tropopause height $z=z_T$, integration of (\ref{eq3.15}) yields  
\begin{equation}                                   
      \bar{\rho}_a \overline{w'{\mathcal S}'} 
    + \bar{\rho}_a \overline{w_\infty {\mathcal S}_r} 
    = \int_{z}^{z_T} \left(\frac{\partial\bar{\mathcal S}}{\partial t} 
                  + \bar{u}\frac{\partial\bar{\mathcal S}}{\partial x}  
                  + \bar{v}\frac{\partial\bar{\mathcal S}}{\partial y}
                  + \bar{w}\frac{\partial\bar{\mathcal S}}{\partial z}     
                  - \frac{\bar{Q}_\sigma}{\bar{\rho}_a} \right)\bar{\rho}_a(z') dz'.  
\label{eq3.16}
\end{equation}
Equation (\ref{eq3.16}) should be compared to (\ref{eq2.13}), which can be considered 
an approximation of (\ref{eq3.16}) in the sense that it neglects vertical transport of 
entropy by precipitation and 
that it uses moist static energy instead of the moist entropy ${\mathcal S}$.  The fluxes 
on the left hand sides of (\ref{eq2.13}) and (\ref{eq3.16}) are often considered the 
primary results of the large-scale heat and moisture budget analysis. 
With the aid of a cloud model, these moist 
static energy or moist entropy fluxes can be interpreted in terms of cloud
mass fluxes and the thermodynamic properties of the air inside the clouds. 
Note that, while the large-scale terms in (\ref{eq2.11}) and (\ref{eq2.13}) are determined entirely from
radiosonde observations, the large-scale terms in (\ref{eq3.13})--(\ref{eq3.15}) require more than just
radiosonde observations. Storage and advection of airborne cloud condensate (which may
be important at upper levels when cirrus occurs) and precipitation are included, as is
the vertical entropy flux by precipitation. 

     The addition of (\ref{eq3.13}) and (\ref{eq3.14}) yields  
\begin{equation}                                  
                  \frac{\partial(\bar{q}_m+\bar{q}_r)}{\partial t} 
         + \bar{u}\frac{\partial(\bar{q}_m+\bar{q}_r)}{\partial x}  
         + \bar{v}\frac{\partial(\bar{q}_m+\bar{q}_r)}{\partial y}
         + \bar{w}\frac{\partial(\bar{q}_m+\bar{q}_r)}{\partial z} 
         =-\frac{\partial(\bar{\rho}_a \overline{w'q'_m}
	                 +\bar{\rho}_a \overline{w'q'_r}
			 +\bar{\rho}_a \overline{w_\infty q_r})}
	        {\bar{\rho}_a\,\partial z}.   
\label{eq3.17}
\end{equation}
Assuming that $\overline{w'q'_m}+\overline{w'q'_r}+\overline{w_\infty q_r}$ vanishes 
at the tropopause height $z=z_T$, integration of (\ref{eq3.17}) yields  
\begin{equation}                                   
    \mathcal{P}_o = \mathcal{E}_o - \int_{0}^{z_T}\left(\frac{\partial(\bar{q}_m+\bar{q}_r)}{\partial t} 
         + \bar{u}\frac{\partial(\bar{q}_m+\bar{q}_r)}{\partial x}  
         + \bar{v}\frac{\partial(\bar{q}_m+\bar{q}_r)}{\partial y}
         + \bar{w}\frac{\partial(\bar{q}_m+\bar{q}_r)}{\partial z}\right) \bar{\rho}_a(z) dz,  
\label{eq3.18}
\end{equation}
where $\mathcal{E}_o=(\bar{\rho}_a\overline{w'q'_m})_o$ is the surface flux of airborne 
moisture and $\mathcal{P}_o=-(\bar{\rho}_a\overline{w'q_r'}+\bar{\rho}_a \overline{w_\infty q_r})_o$ 
is the surface precipitation.\footnote{Note that $\mathcal{E}_o=(\bar{\rho}_a\overline{w'q'_m})_o$ 
and $E_o=(\bar{\rho}_a\overline{w'q'_v})_o$ are identical unless there is cloud (or fog) 
at the surface. The difference between $\mathcal{P}_o$ and $P_o$ is more subtle because 
the improved diagnostic model explicitly tracks the falling precipitation while the standard 
model does not.} 
Equation (\ref{eq3.18}) should be compared to (\ref{eq2.14}), 
which can be considered an approximation to (\ref{eq3.18}). While (\ref{eq2.14}) 
diagnoses rainfall based only on water vapor data, (\ref{eq3.18}) requires 
both water vapor and condensed water data, i.e., the concept of an 
apparent moisture sink is more general. During periods of
high fractional cloudiness this additional information could improve
the budget analyses of rainfall. For selected satellite overpasses when vertical 
profiles of airborne condensed water are available, one could diagnose 
rainfall using both (\ref{eq2.14}) and (\ref{eq3.18}) and compare these analyses with the 
satellite derived rainfall product. By doing such comparisons for a variety
of convective situations, one could learn how different convective conditions 
affect budget-derived rainfall analyses. 

     In summary, we have developed a generalized diagnostic budget method based 
on a more accurate model of the moist atmosphere. Evaluation of budgets using these
more precise techniques is contingent on the availability of certain data products, 
namely, vertical profiles of airborne condensed water and precipitation from 
ground-based and/or satellite observations. This raises the possibility of 
reexamining data from previous field programs to understand what new physical 
processes might be revealed by the more accurate budget methods. Finally we note 
that the symbols $Q_1$ and $Q_2$ and the terms ``apparent heat source" and ``apparent 
moisture sink" have not been used in section 3. The primary reason for this is that, 
while the prognostic thermodynamic principle in section 2 involves enthalpy 
(the $c_{pa}T$ part of the dry static energy),  
the prognostic thermodynamic principle used here in section 3 involves the moist 
entropy (e.g., equations (\ref{eq3.4}), (\ref{eq3.12}), (\ref{eq3.15})) rather 
than enthalpy. While it is possible to convert from the entropy formulation to the 
enthalpy formulation, the analysis is somewhat tedious. Thus, while there is a difference 
in physical detail between section 2 and section 3, there is also a difference in 
focus, with section 3 focusing on the ``apparent entropy source" and section 2 
focusing on the ``apparent heat source."


\section{Nonhydrostatic model of the moist atmosphere}    
   
     In addition to the development of more accurate budget diagnostics, we can 
use a  more accurate nonhydrostatic model of moist convection to simulate 
the convective transports observed in the diagnostic studies. The models that 
have been used are the ones described by Ooyama (2001) for squall lines and 
by Hausman et al.~(2006) for hurricanes. One advantage of this approach is that 
the formulation of moist thermodynamics in the diagnostic studies and in the 
numerical models is identical. The numerical models have seven prognostic 
equations---three momentum equations and the four conservation relations 
(\ref{eq3.1})--(\ref{eq3.4}). The diagnostic relations for the models are  
\begin{equation}                                         
                  \rho = \rho_a + \rho_m + \rho_r,               
\label{eq4.1}
\end{equation}
\begin{equation}                                         
           S_2(\rho_a,\rho_m+\rho_r,T_2) = \sigma,               
\label{eq4.2}
\end{equation}
\begin{equation}                                         
                                \sigma_r = \rho_r C(T_2),           
\label{eq4.3}
\end{equation}
\begin{equation}                                         
                  S_1(\rho_a,\rho_m,T_1) = \sigma - \sigma_r,          
\label{eq4.4}
\end{equation}
\begin{equation}                                         
                       T = \max(T_1,T_2),                             
\label{eq4.5}
\end{equation}
\begin{equation}                                         
                     p_a = \rho_a R_a T,                             
\label{eq4.6}
\end{equation}
\begin{equation}                                         
   \begin{cases}
    \rho_v = \rho_m, \quad \qquad  \rho_c = 0,\quad\qquad p_v = \rho_v R_v T, 
               & {\rm if}\quad  T=T_1>T_2 \quad \text{(absence of cloud condensate)}, \\  
    \rho_v = \rho_v^*(T), \quad \rho_c = \rho_m-\rho_v, \quad p_v = E(T),
               & {\rm if}\quad  T=T_2>T_1 \quad \text{(saturated vapor)}, 
   \end{cases}
\label{eq4.7}
\end{equation}
\begin{equation}                                         
                       p = p_a + p_v,       
\label{eq4.8}
\end{equation}
which introduce the following additional diagnostic variables: 
the total mass density $\rho$, the thermodynamically possible 
temperatures $T_1$ and $T_2$, the actual temperature $T$, the partial 
pressure of dry air $p_a$, the partial pressure of water vapor $p_v$, 
and the total pressure $p$. The functions $S_1(\rho_a,\rho_m,T)$ and 
$S_2(\rho_a,\rho_m,T)$ are given by 
\begin{equation}                                         
      S_1(\rho_a,\rho_m,T) = \rho_a s_a(\rho_a,T) + \rho_m s_m^{(1)}(\rho_m,T),  
\label{eq4.9}
\end{equation}
\begin{equation}                                         
      S_2(\rho_a,\rho_m,T) = \rho_a s_a(\rho_a,T) + \rho_m s_m^{(2)}(\rho_m,T),  
\label{eq4.10}
\end{equation}
where the specific entropy of dry air is 
\begin{equation}                                         
      s_a(\rho_a,T) = c_{va}\ln\left(\frac{T}{T_0}\right) 
                    - R_a   \ln\left(\frac{\rho_a}{\rho_{a0}}\right), 
\label{eq4.11}
\end{equation}
the specific entropy of airborne moisture in state 1 (absence of cloud condensate, 
so that $\rho_v=\rho_m$ and $\rho_c=0$) is 
\begin{equation}                                         
      s_m^{(1)}(\rho_m,T) = c_{vv}\ln\left(\frac{T}{T_0}\right) 
                          - R_v   \ln\left(\frac{\rho_m}{\rho_{v0}^*}\right) 
                          + \frac{L(T_0)}{T_0},    
\label{eq4.12}
\end{equation}
and the specific entropy of airborne moisture in state 2 (saturated vapor, so that 
$\rho_v=\rho_v^*(T)$ and $\rho_c=\rho_m-\rho_v^*(T)$) is 
\begin{equation}                                         
      s_m^{(2)}(\rho_m,T) = C(T) + \frac{D(T)}{\rho_m}
                          = C(T) + \frac{\rho_v^*(T)}{\rho_m}\frac{L(T)}{T},        
\label{eq4.13}
\end{equation} 
with 
\begin{equation}                                         
      C(T) = c_{vv}\ln\left(\frac{T}{T_0}\right) 
           - R_v   \ln\left(\frac{\rho_v^*(T)}{\rho_{v0}^*}\right) 
           + \frac{L(T_0)}{T_0} - \frac{L(T)}{T}        
\label{eq4.14}
\end{equation} 
denoting the entropy of a unit mass of condensate at temperature $T$ (note that 
$C(T_0)=0$), and  
\begin{equation}                                         
      D(T) = \frac{dE(T)}{dT} = \rho_v^*(T) \frac{L(T)}{T}        
\label{eq4.15}
\end{equation} 
denoting the gain of entropy per unit volume by evaporating a sufficient amount 
of water, $\rho_v^*(T)$, to saturate the volume at temperature $T$.

\begin{figure}[tbp]                       
\centering
\includegraphics[width=4.5in]{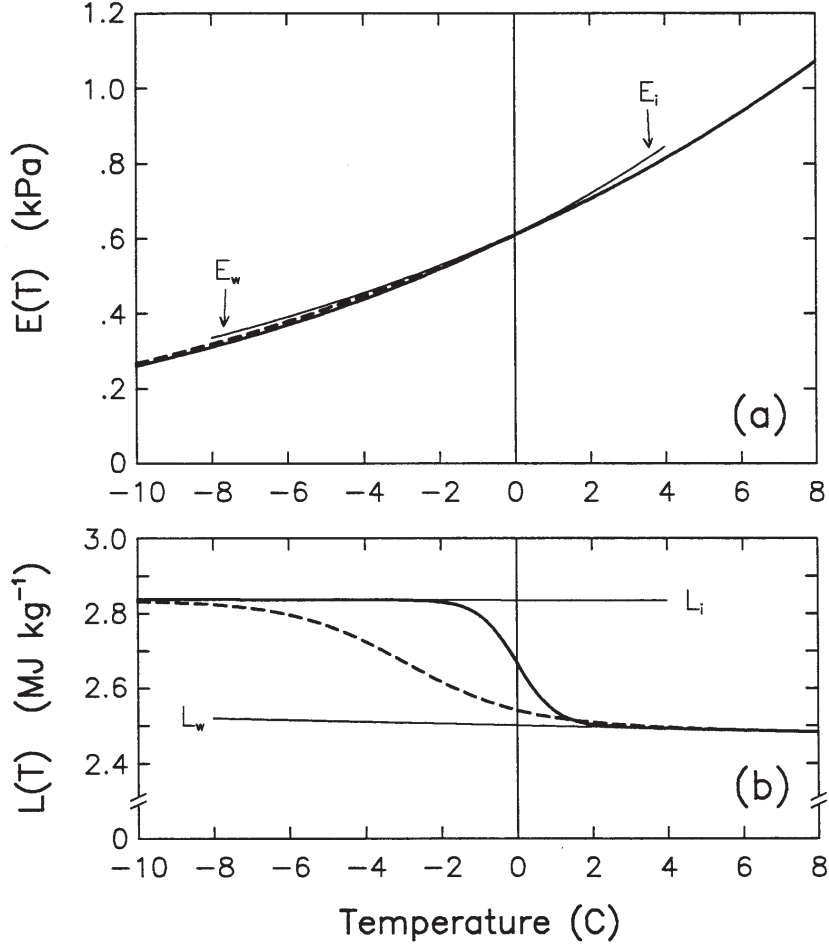}
\caption{The top panel shows the saturation vapor pressure over ice, 
$E_i(T)$, and the saturation vapor pressure over water, $E_w(T)$, with 
$E_w>E_i$ for colder temperatures and $E_i>E_w$ for warmer temperatures.   
These empirical curves are based on laboratory results. They have 
been merged into a single synthetic profile with a sharp transition at 
$0^\circ$C (thick solid curve) and into a single synthetic profile with 
a gradual transition centered at $-3^\circ$C (dashed curve). The $E_w$ 
curve and the two synthetic curves are so close that they are indistinguishable 
on the right side of the upper panel.  The lower panel shows the corresponding 
$L(T)$ curves, which are calculated from the Clausius-Clapeyron equation in the 
form $L(T)=R_v T^2 [d\ln E(T)/dT]$. The difference between the asymptotic values 
$L_i$ and $L_w$ is the latent heat of fusion. Adapted from Ooyama (1990).} 
\end{figure}

     The thermodynamic diagnosis at each spatial point can be interpreted conceptually 
as the following input/output process: 
  $$  \text{Input} \left\{\rho_a,\rho_m,\rho_r,\sigma\right\} \Rightarrow 
      \text{Output}\left\{\rho,T_1,T_2,\sigma_r,T,\rho_v,\rho_c,p_a,p_v,p\right\}.    $$ 
Specifically, starting with the values of the prognostic variables 
$\rho_a,\rho_m,\rho_r,\sigma$, the thermodynamic diagnosis proceeds in the order 
given by (\ref{eq4.1})--(\ref{eq4.8}), i.e., determination of the total density $\rho$ from (\ref{eq4.1}), 
determination of the thermodynamically possible temperature $T_2$ from (\ref{eq4.2}), 
the entropy density of precipitation from (\ref{eq4.3}), the thermodynamically possible 
temperature $T_1$ from (\ref{eq4.4}), the actual temperature $T$ of the gaseous 
matter\footnote{Note that 
there is a physical situation in which not all the matter in a model grid volume has 
the same temperature. This occurs when precipitation is falling through cloud-free 
air. All the gaseous matter has temperature equal to $T_1$, while the precipitation 
has temperature $T_2$, which is lower than $T_1$. This situation is accurately modeled 
by the diagnostic analysis (\ref{eq4.1})--(\ref{eq4.8}) and is consistent with our common experience of 
cold raindrops falling on our skin.} from (\ref{eq4.5}), the partial pressure of 
dry air from (\ref{eq4.6}), the water vapor density $\rho_v$, cloud condensate density $\rho_c$, 
and water vapor partial pressure $p_v$ from the appropriate alternative in (\ref{eq4.7}), and 
the total pressure $p$ from (\ref{eq4.8}). Note that all the other required thermodynamic 
functions, such as $C(T)$, $D(T)$, $\rho_v^*(T)$, etc., can be determined from $E(T)$, once 
it is specified. If $E(T)$ is specified as $E_w(T)$ for all $T$, then the effects 
of the latent heat of fusion are not included. If a synthesized $E(T)$ is obtained from 
$E_w(T)$ and $E_i(T)$, then the effects of the latent heat of fusion are included.  

     To summarize, the procedure for advancing from one time level to the 
next consists of computing new values of the prognostic variables 
$\rho_a,\rho_m,\rho_r,\sigma,u,v,w$ from (\ref{eq3.1})--(\ref{eq3.4}) and the three momentum 
equations. The diagnostic variables required for the prognostic stage are 
determined by sequential evaluation of (\ref{eq4.1})--(\ref{eq4.8}). Since they are not essential 
to our discussion here, we have omitted the  
parameterization formulas for the terminal fall velocity, $w_\infty$, and 
the source terms $Q_r$ and $Q_\sigma$.\footnote{We simply note that 
the terminal fall velocity $w_\infty$ of precipitating water or ice 
is parameterized in terms of $\rho_r$, $\rho_a$ and $T$, with a slow 
terminal velocity for ice when $T$
is below the freezing point and a larger terminal velocity for rain when $T$ is above
the freezing point, and that the rate of conversion to precipitation $Q_r$ 
is usually parameterized as the sum of the rates of autoconversion (dependent on
$\rho_c$ and $\rho_a$), collection (dependent on $\rho_c$, $\rho_r$ and $\rho_a$), 
and evaporation (dependent on $\rho_v$, $\rho_r$ and $T$).}

\begin{figure}[tbp]                       
\centering
\includegraphics[width=6.0in]{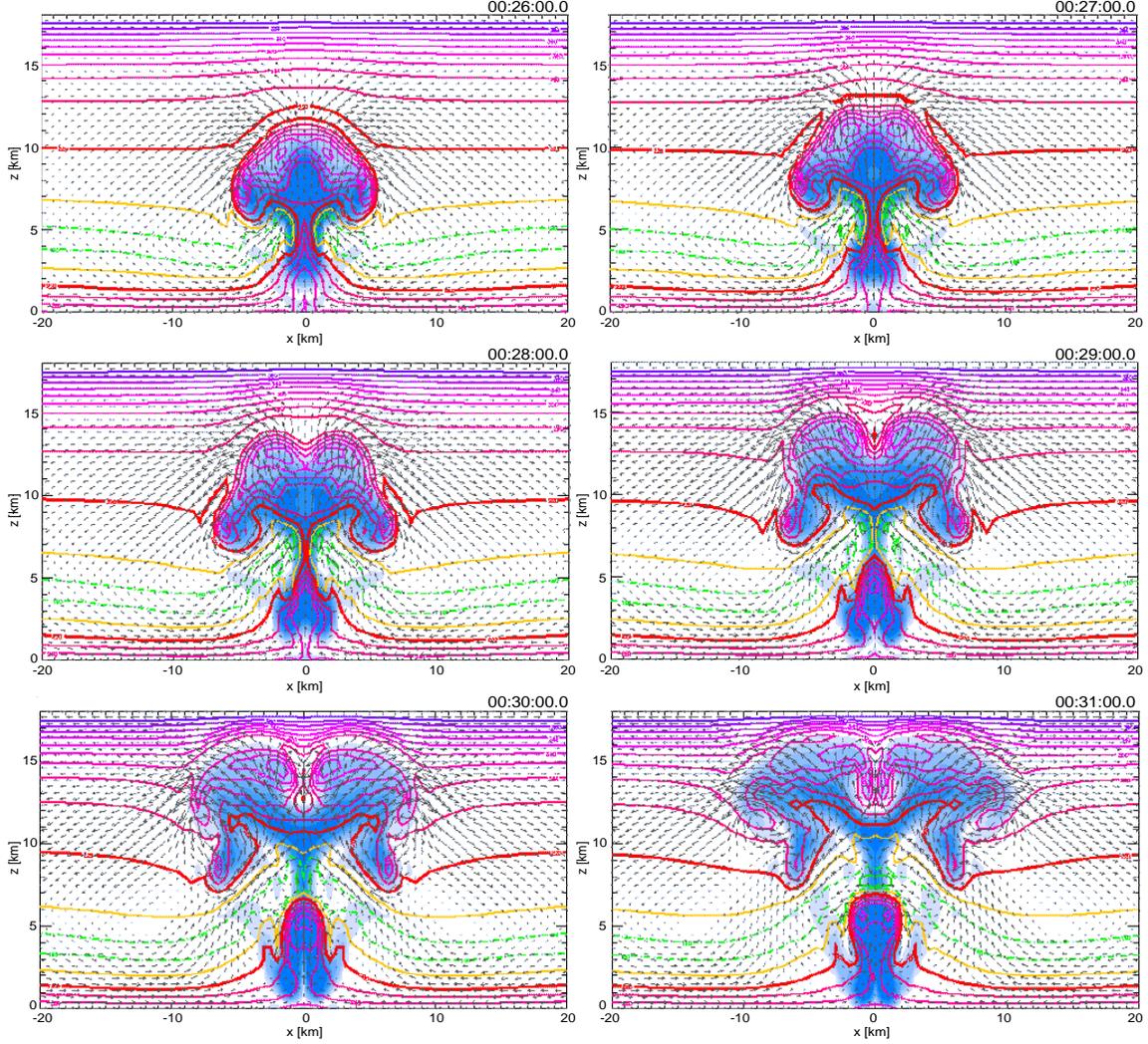}
\caption{Development of a convective cell at one minute intervals using
the nonhydrostatic moist model developed by Ooyama (1990, 2001). Model time is
specified in the upper right of each panel, contour lines indicate the specific
entropy of moist air, and blue shading indicates cloud boundaries and regions
of precipitation. Note the transition from a plume of high $\theta_e$ air to a
bubble structure below 5 km. Adapted from Garcia (1999).} 
\end{figure}

    It is interesting to note how the variation of $L$ with temperature is 
incorporated into this model. The idea is easily understood by reference to Fig.~1, 
the top panel of which shows the saturation vapor pressure over ice, $E_i(T)$, and 
the saturation vapor pressure over water, $E_w(T)$. Also shown are two merged $E(T)$ 
profiles, one with a sharp transition centered at $0^\circ$C (solid curve) and one 
with a gradual transition centered at $-3^\circ$C (dashed curve). The lower panel 
of Fig.~1 shows the two corresponding $L(T)$ curves, which have been calculated from 
the Clausius-Clapeyron equation in the form $L(T)=R_v T^2 [d\ln E(T)/dT]$. The choice 
of the solid curve would produce model results with a sharp melting layer, while 
choice of the dashed curve would produce a melting layer several times as thick. 

     At present it is not feasible to numerically 
integrate such a nonhydrostatic ``full physics" model over the whole globe 
with 1--2 km resolution. However, it is possible to perform such high 
resolution integrations over the limited area of a single tropical disturbance. 
Such integrations advance the art of tropical modeling to a new level that 
involves much less physical parameterization.  An example of a high resolution 
simulation using this model is shown in Fig.~2, which depicts 
the development of convection in an environment at rest (Garcia 1999).  
For this simulation, the domain consists of 5 nested grids over a 
total horizontal distance of 1152 km.  The finest 
(central) grid in this domain spans 96 km with a horizontal grid spacing of 
0.5 km.  The height of the domain is 18 km, with a constant vertical  
resolution of 750 m on all grids.  The time integration for this 
simulation uses a semi-implicit formulation of the 
leapfrog method with a time step of 2.5 seconds on the finest grid.  The 
initial background for this simulation consists of a  
mean hurricane season sounding which has been modified slightly to 
include higher relative humidity in the lower levels (surface to 
approximately 700 hPa) in order to better represent disturbed tropical 
conditions.  The initial perturbation employed here was a thermal anomaly 
of $T'=+3$K, centered at $x=0$, $z=0$ and with an overall width of 32 
km.  This figure depicts the development of the resulting convective 
cell at 1 minute intervals between 26 and 31 minutes of a 1 hour  
simulation. The line contours in Fig.~2 represent the dry-air-specific entropy 
of moist air as calculated by ${\mathcal S} = \sigma/\rho_a$. The contour interval 
is 10 J kg$^{-1}$ K$^{-1}$. The dry-air-specific entropy of moist air can 
also be interpreted in terms of equivalent potential temperature $\theta_e$ 
through the definition  
  $$ \theta_e = T_0 e^{{\mathcal S}/c_{pa}}  \quad \Longleftrightarrow \quad 
       {\mathcal S} = c_{pa} \ln(\theta_e/T_0).                  $$ 
Thus, ${\mathcal S}=$ 200 J kg$^{-1}$ K$^{-1}$ (the two neighboring green 
contours near $z\approx 4$ km in the cloud environment) correspond to 
$\theta_e = 333$K, while the ${\mathcal S} =$ 250 J kg$^{-1}$ K$^{-1}$ contour 
near the surface in most of these panels corresponds approximately to 
$\theta_e = 348$K.  The blue shaded regions indicate the cloud water and 
precipitation content of the air:  the lightest blue corresponds to cloud 
water in any concentration, while the darker blue shades represent 
precipitation content at intervals of $0.5$ g kg$^{-1}$. An interesting 
feature of Fig.~2 is the way in which the high $\theta_e$ contours 
(or, equivalently, the high $h$ contours) near the surface are drawn up 
in the convective plume during the first two minutes. These contours then 
close off and a bubble of high $\theta_e$ air is carried upward. Another 
interesting feature is the cap instability that develops near $z\approx 13$ km, 
above the rising convective plume. The $\bar{\rho}_a\overline{w'{\mathcal S}'}$ 
flux in (\ref{eq3.16}) can be considered to be the result of a field of such convective 
clouds in various stages of their life cycles. 

   Numerous simulations have been performed with this model. These include 
simulations of tropical squall line systems (Ooyama 2001), and simulations 
of tropical cyclone development (Hausman et al.~2006) using an axisymmetric 
version of the model. In this regard it is interesting to note that such 
model simulations can lead to a better 
understanding of CAPE (or the cloud work function, which is a more general 
concept) and its generation by large-scale processes. To see why further understanding
of convective energetics is needed, consider the following question. 
How do we define CAPE? The theory just described gives us at least eight ways! We
can define a reversible adiabat (RA) with all condensate carried along with the
parcel, and a pseudoadiabat (PA) with all condensate removed. Each of these can be
defined with ice (wi) or with no ice (ni). Thus, we have four temperature profiles 
and four virtual temperature profiles for RA.wi, PA.wi, RA.ni, PA.ni. This gives eight
ways of computing CAPE. Of course, we can also define CAPE 
in a myriad of ways which lie somewhere between RA and PA.

\section{Concluding remarks}

     The main focus of our discussion has been the development of new, 
more accurate methods for the diagnostic analysis of heat and moisture 
budgets. The physical model which serves as the basis for this discussion   
is the nonhydrostatic moist model, consisting of (\ref{eq3.1})--(\ref{eq3.4}) and three 
momentum equations. In this model, 
pressure is not used as one of the prognostic variables, since it is not a
conservative property and its use as a prognostic variable would lead to an
approximate treatment of moist thermodynamics. 
Rather, the prognostic variable for the thermodynamic state is $\sigma$, 
the entropy of moist air per unit volume, with temperature and total pressure (the
sum of the partial pressures of dry air and water vapor) determined
diagnostically. There are several unique aspects of this model which are worth noting. 
(1) The model dynamics are exact in the sense that there is no hydrostatic approximation.
(2) The connection between dynamics and thermodynamics is through the gradient 
of pressure, which includes the partial
pressures of dry air and water vapor. (3) The first law of thermodynamics is expressed
in terms of $\sigma$, the entropy density of moist air; all the usual approximations
associated with moist thermodynamics are thereby avoided. (4) There is no cumulus
parameterization; the frontier of empiricism is pushed back to the
microphysical parameterization of the precipitation process through $w_\infty$ 
and $Q_r$. (5) The model is modular in the sense that ice can be included 
by specifying $E(T)$ to be synthesized from the saturation formulas over water and ice. 

     In spite of the above unique aspects, the model thermodynamics should not 
be regarded as ``exact." For example, the effects of the latent heat of fusion 
are always recognized in the same temperature interval (see Fig.~1), whether 
freezing is occurring on ascent or melting is occurring on descent. Hence, highly 
supercooled water in intense thunderstorm updrafts cannot be simulated. In any 
event, the concepts presented here open up the possibility of coordinated 
modeling and observational budget studies based on the same moist thermodynamic 
principles. At present, this research path remains largely unexplored.

\acknowledgments
We would like to thank Rick Taft for his help in preparing this paper. 
The authors' research has been supported by the National Science Foundation 
under Grants AGS-1360237, AGS-1546610, and AGS-1601623.  

\appendix{List of Key Symbols for the Nonhydrostatic Model}

\smallskip
\noindent
\begin{tabular}{p{1em}p{1.65in}p{4.2in}}
\multicolumn{3}{l}{Constants:}\\[0.5ex]
 & $R_a$    & gas constant of dry air  \\
 & $R_v$    & gas constant of water vapor \\ 
 & $c_{va}$ & specific heat of dry air at constant volume \\ 
 & $c_{vv}$ & specific heat of water vapor at constant volume\\ 
 & $c_{pa}=c_{va}+R_a$ & specific heat of dry air at constant pressure \\ 
 & $c_{pv}=c_{vv}+R_v$ & specific heat of water vapor at constant volume\\ 
 & $\kappa = R_a/c_{pa}$ & \\    
 & $p_{a0}$              & reference pressure, 100 kPa  \\ 
 & $T_0$                 & reference temperature, 273.15 K \\
 & $\rho_{a0}=p_{a0}/(R_aT_0)$ & reference density for dry air  \\
 & $\rho_{v0}^*=\rho_v^*(T_0)$ & mass density of saturated vapor at $T_0$ \\[2ex]
\multicolumn{3}{l}{Velocities (m s$^{-1}$):}\\[0.5ex]
 & $(u,v,w)$ & velocity of dry air and airborne moisture \\
 & $ w_r$    & vertical velocity of precipitation (relative to earth) \\ 
 & $ w_\infty = w_r - w$ 
    & terminal velocity of precipitation (relative to dry air and airborne
      moisture) \\[2ex]
\multicolumn{3}{l}{Pressures (Pa):}\\[0.5ex]
 & $p_a$ & partial pressure of dry air  \\ 
 & $p_v$ & partial pressure of water vapor  \\
 & $p=p_a + p_v$ & total pressure
\end{tabular}

\noindent
\begin{tabular}{p{1em}p{1.65in}p{4.2in}}
\multicolumn{3}{l}{Temperatures (K):}\\[0.5ex]
 & $T_1$
    & thermodynamically possible temperature (absence of cloud condensate) \\
 & $T_2$
    & thermodynamically possible temperature (saturated vapor) \\
 & $T=\text{max}(T_1,T_2)$ & temperature \\
 & $\theta_e=T_0 e^{{\mathcal S}/c_{pa}}$
    & equivalent potential temperature \\[2ex]
\multicolumn{3}{l}{Mass Densities (kg m$^{-3}$) and Mixing Ratios:}\\[0.5ex]
 & $\rho_a$ & mass density of dry air \\
 & $\rho_v$ & mass density of water vapor \\
 & $\rho_c$ & mass density of airborne condensate (water droplets or ice
              crystals) \\
 & $\rho_r$ & mass density of precipitating water substance (liquid or ice) \\
 & $\rho_m = \rho_v + \rho_c$ 
    & mass density of airborne moisture (vapor plus airborne condensate) \\
 & $\rho   = \rho_a + \rho_m + \rho_r$
    & total mass density (dry air plus airborne moisture plus precipitation) \\
 & $q_v=\rho_v/\rho_a$ & mixing ratio of water vapor \\
 & $q_m=\rho_m/\rho_a$ & mixing ratio of airborne moisture \\
 & $q_r=\rho_r/\rho_a$ & mixing ratio of precipitation \\[2ex]
\multicolumn{3}{l}{Specific Entropies (J kg$^{-1}$ K$^{-1}$):}\\[0.5ex]
 & $s_a$ & specific entropy of dry air \\ 
 & $s_m^{(1)}$ & specific entropy of airborne moisture in state 1 (absence of cloud condensate) \\ 
 & $s_m^{(2)}$ & specific entropy of airborne moisture in state 2 (saturated vapor) \\ 
 & $s_m$ & specific entropy of airborne moisture (vapor and cloud) \\ 
 & $s_r$ & specific entropy of condensed water (cloud or precipitation) \\
 & ${\mathcal S}  =\sigma  /\rho_a$ & dry-air-specific entropy of moist air \\ 
 & ${\mathcal S}_r=\sigma_r/\rho_a$ & dry-air-specific entropy of precipitation \\[2ex]
\multicolumn{3}{l}{Entropy Densities (J m$^{-3}$ K$^{-1}$):}\\[0.5ex]
 & $\sigma_a = \rho_a s_a$ & entropy density of dry air \\ 
 & $\sigma_m = \rho_m s_m$ & entropy density of airborne water substance \\ 
 & $\sigma_r = \rho_r s_r$ & entropy density of precipitating water substance \\ 
 & $\sigma   = \sigma_a + \sigma_m + \sigma_r$ & total entropy density \\
 & $S_1(\rho_a,\rho_m,T)$ & entropy density function of dry air and airborne moisture (absence of cloud \\
 &                        & \hspace{1em} condensate) \\ 
 & $S_2(\rho_a,\rho_m,T)$ & entropy density function of dry air and airborne moisture (saturated vapor) \\[2ex] 
\multicolumn{3}{l}{Defined Functions of Temperature:}\\[0.5ex]
 & $L(T)=R_v T^2(d\ln E(T)/dT)$
    & latent heat of vaporization, computed from the Clausius-Clapeyron
      equation \\ 
 & $L(T)/T$
    & gain of entropy by evaporating a unit mass of water at $T$ \\
 & $C(T)$
    & entropy of a unit mass of condensate at $T$, as measured from the
      reference \\
 &  & \hspace{1em} state $T_0$ \\
 & $D(T)=dE(T)/dT$
    & gain of entropy per unit volume by evaporating a sufficient amount of
      water, \\
 &  & \hspace{1em} $\rho_v^*(T)$, to saturate the volume at $T$ \\
 & $E_w(T)$
   & saturation vapor pressure over water \\ 
 & $E_i(T)$
   & saturation vapor pressure over ice \\ 
 & $E(T)$
    & saturation vapor pressure, which may be synthesized from the saturation \\
 &  & \hspace{1em} vapor pressures over water and ice \\ 
 & $\rho_v^*(T)=E(T)/(R_vT)$ & mass density of saturated vapor
\end{tabular}

\end{document}